\documentclass[aps,prl,preprint,superscriptaddress]{revtex4-1}

\usepackage{hyperref}
\usepackage{graphicx}

\begin{document}

\graphicspath{{}{figures/}}

\title{Direct measurement of time-dependent density-density correlations in a solid through the acoustic analog of the dynamical Casimir effect}

%

\author{M. Trigo}
\email[]{mtrigo@slac.stanford.edu}
\affiliation{Stanford Institute for Materials and Energy Sciences, SLAC National Accelerator Laboratory, Menlo Park, CA 94025, USA}
\affiliation{Stanford PULSE Institute, SLAC National Accelerator Laboratory, Menlo Park, CA 94025, USA}
\author{M. Fuchs}
\altaffiliation{current address: University of Nebraska Lincoln, Lincoln, Nebraska 68588, USA}

\author{J. Chen}
\author{M. P. Jiang}
\author{M. E. Kozina}
\author{G. Ndabashimiye}
\affiliation{Stanford PULSE Institute, SLAC National Accelerator Laboratory, Menlo Park, CA 94025, USA}

\author{M. Cammarata}
\affiliation{Linac Coherent Light Source, SLAC National Accelerator Laboratory, Menlo Park, CA 94025, USA}

\author{G. Chien}
\affiliation{Department of Physics, University of Michigan, Ann Arbor, MI 48109, USA}

\author{S. Fahy}
\affiliation{Tyndall National Institute and Department of Physics, University College, Cork, Ireland}

\author{D. M. Fritz}
\affiliation{Linac Coherent Light Source, SLAC National Accelerator Laboratory, Menlo Park, CA 94025, USA}

\author{K. Gaffney}
\affiliation{Stanford PULSE Institute, SLAC National Accelerator Laboratory, Menlo Park, CA 94025, USA}

\author{S. Ghimire}
\affiliation{Stanford PULSE Institute, SLAC National Accelerator Laboratory, Menlo Park, CA 94025, USA}

\author{A. Higginbotham}
\affiliation{Department of Physics, Clarendon Laboratory, University of Oxford, Parks Road, Oxford OX1 3PU, United Kingdom}

\author{S. L. Johnson}
\affiliation{Physics Department, ETH Zurich, 8093 Zurich, Switzerland}

\author{J. Larsson}
\affiliation{Department of Physics, Lund University, S-22100 Lund, Sweden}

\author{H. Lemke}
\affiliation{Linac Coherent Light Source, SLAC National Accelerator Laboratory, Menlo Park, CA 94025, USA}

\author{A. M. Lindenberg}
\affiliation{Stanford PULSE Institute, SLAC National Accelerator Laboratory, Menlo Park, CA 94025, USA}
\affiliation{Stanford Institute for Materials and Energy Sciences, SLAC National Accelerator Laboratory, Menlo Park, CA 94025, USA}
\affiliation{Department of Materials Science and Engineering, Stanford University, Stanford, CA 94305, USA}

\author{F. Quirin}
\affiliation{Faculty of Physics and Center for Nanointegration Duisburg-Essen (CENIDE), University of Duisburg-Essen 47048, Duisburg, Germany}

\author{K. Sokolowski-Tinten}
\affiliation{Faculty of Physics and Center for Nanointegration Duisburg-Essen (CENIDE), University of Duisburg-Essen 47048, Duisburg, Germany}

\author{C. Uher}
\affiliation{Department of Physics, University of Michigan, Ann Arbor, MI 48109, USA}
\author{J. S. Wark}
\affiliation{Department of Physics, Clarendon Laboratory, University of Oxford, Parks Road, Oxford OX1 3PU, United Kingdom}

\author{D. Zhu}
\affiliation{Linac Coherent Light Source, SLAC National Accelerator Laboratory, Menlo Park, CA 94025, USA}

\author{D. A. Reis}
\affiliation{Stanford Institute for Materials and Energy Sciences, SLAC National Accelerator Laboratory, Menlo Park, CA 94025, USA}
\affiliation{Stanford PULSE Institute, SLAC National Accelerator Laboratory, Menlo Park, CA 94025, USA}
\affiliation{Department of Photon Science and Applied Physics, Stanford University, Stanford, CA 94305, USA}

\date{\today}

\begin{abstract}

\end{abstract}

\pacs{}

\maketitle

The macroscopic characteristics of a solid, such as its thermal, optical or transport properties are determined by the available microscopic states above its lowest energy level. 
These slightly higher quantum states are described by elementary excitations (phonons, magnons, quasi-particles) and dictate the  response of the system under external stimuli.
The spectrum of these excitations, obtained typically from inelastic neutron\cite{brockhouse1955} and x-ray\cite{rueff2010,krisch_book} scattering, is the spatial and temporal Fourier transform of the density-density correlation function (fluctuations) of the system, which dictates how a perturbation propagates in space and time\cite{vanhove1954}.
As frequency-domain measurements do not generally contain phase information\cite{abbamonte2004}, time-domain measurements of these fluctuations could yield a more direct method for investigating the excitations of solids and their interactions both in equilibrium and far-from equilibrium.
Here we show that the diffuse scattering of femtosecond x-ray pulses produced by a free electron laser (FEL) can directly measure these density-density correlations due to lattice vibrations in the time domain. We obtain spectroscopic information of the lattice excitations with unprecedented momentum- and frequency- resolution, without resolving the energy of the outgoing photon, by a Fourier transform of the femtosecond dynamics.
Correlations are created via an acoustic analog of the dynamical Casimir effect\cite{jaskula2012,carusotto2009}, where a femtosecond laser pulse slightly quenches the phonon frequencies, producing pairs of squeezed phonons at momenta $+q$ and $-q$\cite{garrett1997,johnson2009}.
These pairs of phonons manifest as macroscopic, time-dependent coherences in the displacement correlations\cite{glauber1955} that are then probed directly by x-ray scattering. 
Since the time-dependent correlations are preferentially created in regions of strong electron-phonon coupling, the time-resolved approach is natural as a spectroscopic tool of low energy collective excitations in solids, and their microscopic interactions, both in linear response and beyond.


Spatial density fluctuations in nominally periodic media reduce the intensity of the Bragg diffraction peaks and consequently increase the weak diffuse scattering between these peaks, the details of which reflect the amplitudes and spatial frequencies of the fluctuations\cite{Warren_book}. 
When these perturbations are due to lattice vibrations, the intensity of the diffuse scattering at momentum transfer $Q$ is proportional to the density-density correlation $\langle u_{q} u_{-q} \rangle$, where $u_q$ is the phonon amplitude at reduced wavevector $q = Q - K_Q$ and $K_Q$ is the closest reciprocal lattice vector to $Q$\cite{Warren_book,chiang2005}. In typical experiments with long x-ray (or neutron) pulses, the measured diffuse scattering is proportional to the time-integrated mean squared displacements, and spectral information can only be obtained by analyzing the energy of the scattered particles.
Recent advances in FEL sources\cite{emma2010,amann2012} provide sufficient flux and time-resolution to record snapshots of  the scattered x rays on a time-scale short compared to the density fluctuations. 
As we show here in a single-crystal of the prototypical semiconductor germanium, temporal coherences in $\langle u_{q} u_{-q} \rangle(t)$ induced by a sudden softening of the lattice yield the phonon dispersion without having to resolve the scattered photon energy.

\begin{figure*}
\includegraphics[scale=0.65]{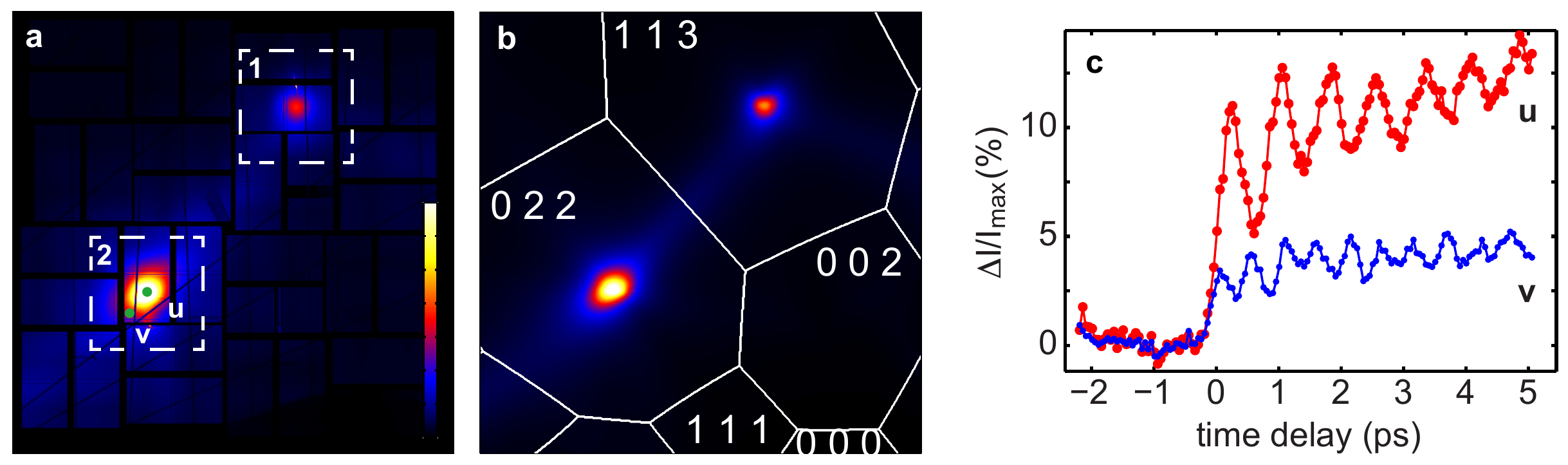}%
\caption{{\bf Femtosecond x-ray diffuse scattering} (a) Static thermal diffuse scattering from (001) Ge in grazing incidence from 10 keV x-ray photons at the LCLS. Dashed boxes are the q-space regions shown in Fig.~\ref{Fig3}.  (b) Calculated equilibrium pattern using a Born model of the forces. White lines indicate the boundaries of the Brillouin zones. Miller indices are also indicated. (c) Representative traces of the normalized change in scattering $\Delta I(t)/I_{\rm max}$ induced by the optical laser as a function of (optical) pump-(x-ray) probe delay. 
\label{Fig1}}
\end{figure*}

Figure \ref{Fig1} (a) shows a portion of the equilibrium x-ray diffuse scattering from a single-crystal of germanium at grazing incidence captured with an area detector (see supplemental material). The bright areas correspond to regions of reciprocal space with low frequency phonons that contribute strongly to the equilibrium diffuse scattering, mostly dominated by thermally populated acoustic phonons~\cite{trigo2010,chiang2005}.
Figure \ref{Fig1} (b) shows the simulated thermal diffuse scattering from a Born model of the forces \cite{Warren_book,trigo2010,chiang2005} (see supplemental information for additional details). The calculated pattern matches the measured diffuse scattering extremely well.  The white solid contours in Fig.~\ref{Fig1} (b) represent the boundaries of the Brillouin zones accessible in this geometry and we have also indicated the respective Miller indices. 
We note that in some specific situations\cite{holt1999}, the phonon dispersion can be obtained from a fit of the calculated image [Fig.~\ref{Fig1} (b)] to the experimental data [Fig.~\ref{Fig1} (a)]. 
Instead, here we show that coherences in the displacement correlations can yield large sections of the phonon dispersion directly from the measurement.

Figure \ref{Fig1} (c) shows the evolution of the change in diffuse scattering intensity $\Delta I(t)/I_{\rm max}$ induced by photoexcitation with a 50 fs infrared laser pulse centered at 800 nm. The two curves show the time traces for the two points labeled ``u'' and ``v'' in Fig. \ref{Fig1} (a), normalized by the maximum of the laser-off image. Photoexcitation induces an overall step-like increase in the scattering whose magnitude depends on momentum position, and oscillations at frequencies in the acoustic phonon range 1 - 3.5 THz. 
The sharpness of the initial step as well as the highest frequency observed $\sim 3.5$~THz were resolution-limited by the timing jitter in the pump-probe delay $\sim $~250~fs\cite{glownia2010}.
As we discuss next, we identify these oscillations as the resulting time-dependent correlations, $\langle u_{q} u_{-q} \rangle(t)$, due to the formation of (squeezed) pairs of phonons from the sudden softening of the lattice forces. 

We begin by considering a sudden quench (softening) in the  harmonic potential driven by excitation of electron-hole pairs by the laser pulse. 
In the specific case of the tetrahedrally bonded semiconductors, photoexcitation from the mainly bonding valence band states to the mainly anti-bonding conduction band is predicted to soften the transverse acoustic (TA) modes\cite{biswas1980,stampfli1990,stampfli1992,zijlstra2008,hillyard2008}, which at high enough excitation density leads to the loss of crystalline order through an instability of the shear (TA) modes. 
In more general terms, the sudden quench of the potential is a consequence of the electron phonon coupling and thus is not restricted to the specifics of these materials.
The evolution of a harmonic oscillator after a sudden quench of the frequency is related to vacuum squeezing, where the variance in the quantum displacements oscillates below the quantum limit for a quarter of the oscillator cycle at the expense of uncertainty in the momentum, as shown for photons\cite{slusher1985,wall1986} and phonons\cite{garrett1997,johnson2009}.
This is intimately related with the dynamical Casimir effect\cite{nation2012} in which a  sudden change in the boundary conditions of the electromagnetic field generates real photons out of the vacuum  quantum fluctuations\cite{wilson2011}. More recently, an acoustic analog of this effect was demonstrated in a Bose-Einstein condensate where a sudden quench of the sound velocity was shown to produce correlated pairs of phonons\cite{jaskula2012}.

Assuming all phonons, with frequencies $\Omega_q$ and oscillator mass $m$, are in the ground state, a sudden change in the frequency $\Omega_q \to \Omega_q'$ at $t=0$ leaves each mode in a squeezed state  
 where the variance in the displacement evolves according to\cite{kiss1994} 
\begin{equation}\label{sqeezed_phonon}
	\langle u_{q} u_{-q} \rangle(t ) = \frac{1}{4 m \Omega_q} \left[( 1  + \beta_q^2) + (1- \beta_q^2)\cos(2 \Omega_q' t) \right],
\end{equation}
where $\beta_q = \Omega_q/\Omega_q'>1$  for a sudden softening. At finite temperature, Eq.~(\ref{sqeezed_phonon}) contains an additional thermal factor\cite{carusotto2009}.
This expression describes the generation of correlated pairs of phonons at $q$ and $-q$\cite{carusotto2009}. This mechanism is formally analogous to (spontaneous) parametric down-conversion.
For simplicity we have assumed that the equilibrium position of the oscillator is unchanged in the excited state, i.e. there is no coherent displacement and thus $\langle u(t) \rangle =0 $\cite{garrett1997}, but this is not a requirement. 
The momentum of the pump photons is small compared to the Brillouin zone dimensions, thus conservation of momentum dictates that the generated pairs of phonons have equal and opposite momenta. 
In our case, the softening occurs for all $q$ and is expected to be particularly strong at the Brillouin zone boundary\cite{zijlstra2008,hillyard2008}, but due to the current time-resolution limit, the phonon frequencies at the zone boundaries were not resolved. 

\begin{figure}
\includegraphics[scale=0.4]{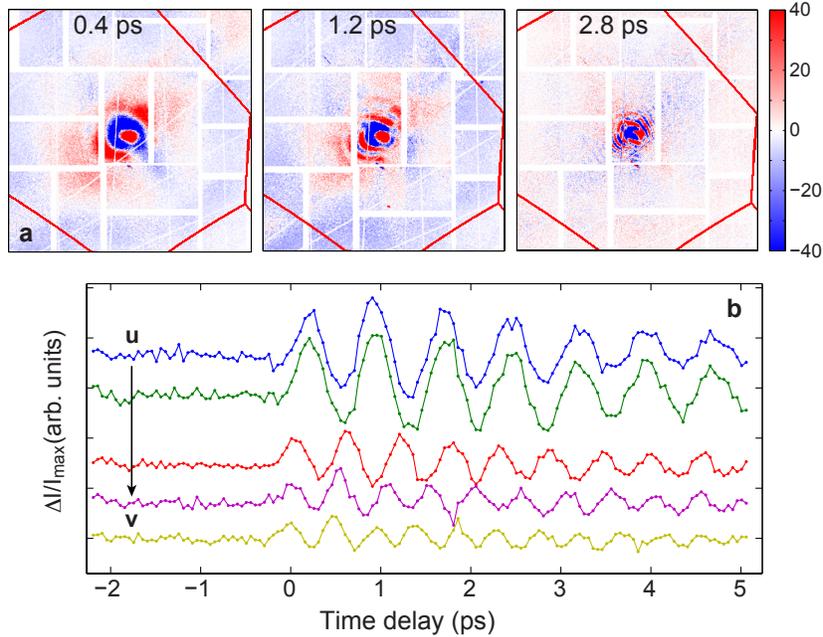}%
\caption{{\bf Coherence in the density-density correlations.} (a) Representative frames of the oscillatory component of $\Delta I/I_{\rm max}$  after background subtraction. (b) Time dependence of the subtracted data at a few reduced wavevector locations between $u =$\,[-0.09   -0.03   -0.08] and  $v = $\,[-0.08    0.13   -0.27]  (r.l.u.) in Fig.~\ref{Fig1} (a). These curves have been displaced vertically for clarity.
\label{Fig2}}
\end{figure}

The x-ray diffuse scattering intensity is directly proportional to the projection of the time-dependent correlations [Eq.~(\ref{sqeezed_phonon})] along $Q$\cite{Warren_book}. 
Accordingly, the intensity oscillates at $2 \Omega'$, with an amplitude proportional to $1-\beta^2_q$. In the limit of low-density excitation, the frequencies will approximate the equilibrium values, and thus the Fourier transform of the oscillatory component should give the phonon dispersion. In our case, $|1-\beta^2_q| \approx 0.05$ and thus $\Omega_q/\Omega'_q \approx 1.025$, which is close to the frequency resolution limit of $\sim 0.1$~THz given by the finite time window in these data.
Consistent with bond softening the mean square displacements (and thus the scattering) increases during the first quarter cycle, as seen in Fig.~\ref{Fig1} (c).

For better sensitivity to the oscillatory signal we filtered the slowly varying background from  the raw data (supplemental information and movies).
Figure \ref{Fig2} (a) shows representative frames of the obtained oscillatory component. The plots in Fig.~\ref{Fig2} show an enlarged view of the (022) Brillouin zone. The data for zone (113) shows similar results. The blue (red) regions in this figure represent an increase (decrease) in the intensity relative to the subtracted average. 
The fringes in $q$-space seen here originate from phonons with different frequencies across reciprocal space, which have phase coherence due to the sudden frequency softening.
The traces in Fig. \ref{Fig2} (b) show some of these oscillations for a few wavevectors along the $u - v$ segment in Fig.~\ref{Fig1} (a). 

\begin{figure}
\includegraphics[scale=0.4]{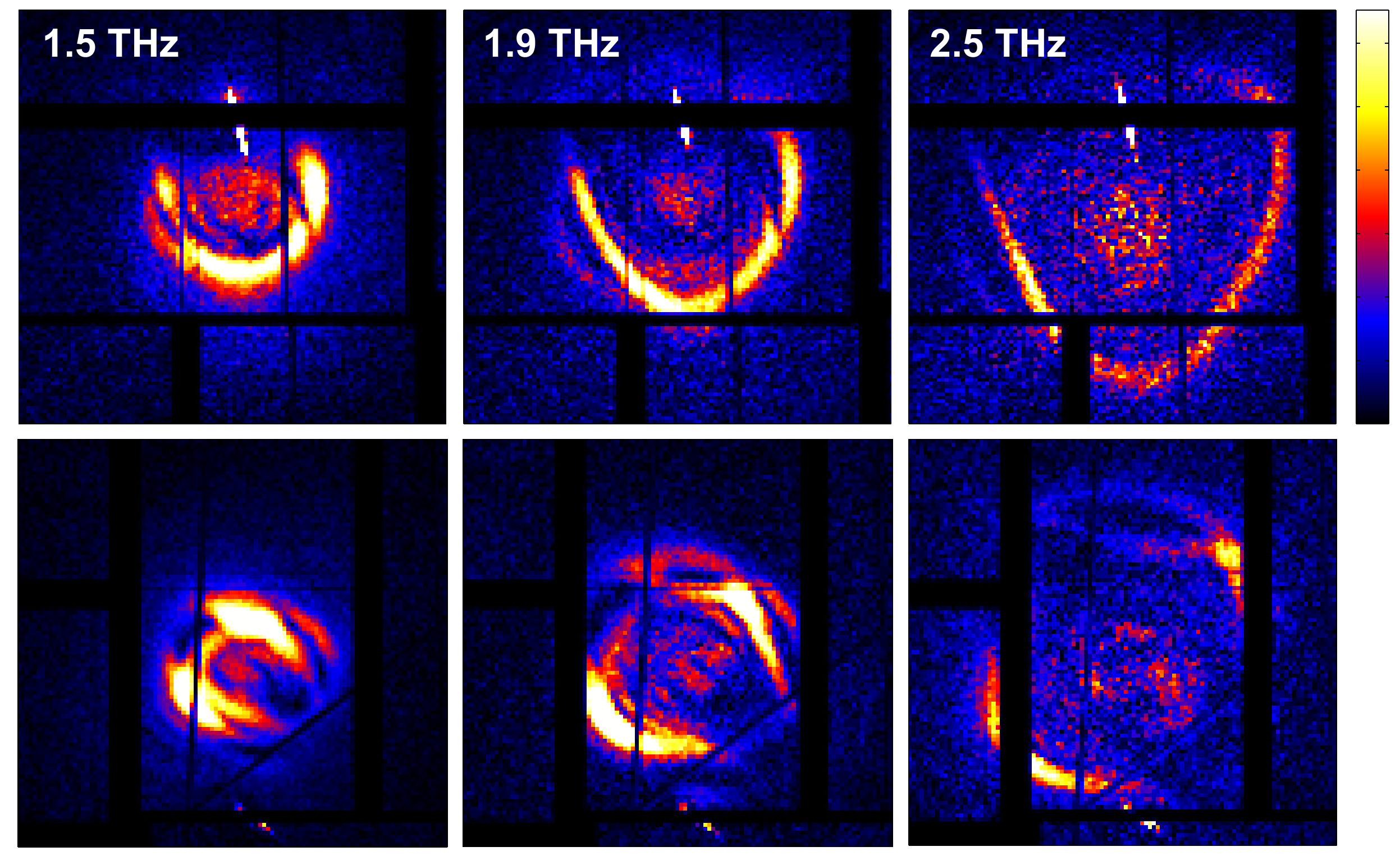}%
\caption{{\bf Equal energy phonon momentum distribution.} Magnitude of the time-Fourier transform at representative frequencies of the background-subtracted data. (top and bottom panels) zoomed view of the region of $q-$space labeled ``1'' and ``2'' in Fig.~\ref{Fig1}, respectively.
\label{Fig3}}
\end{figure}

\begin{figure}
\includegraphics[scale=0.6]{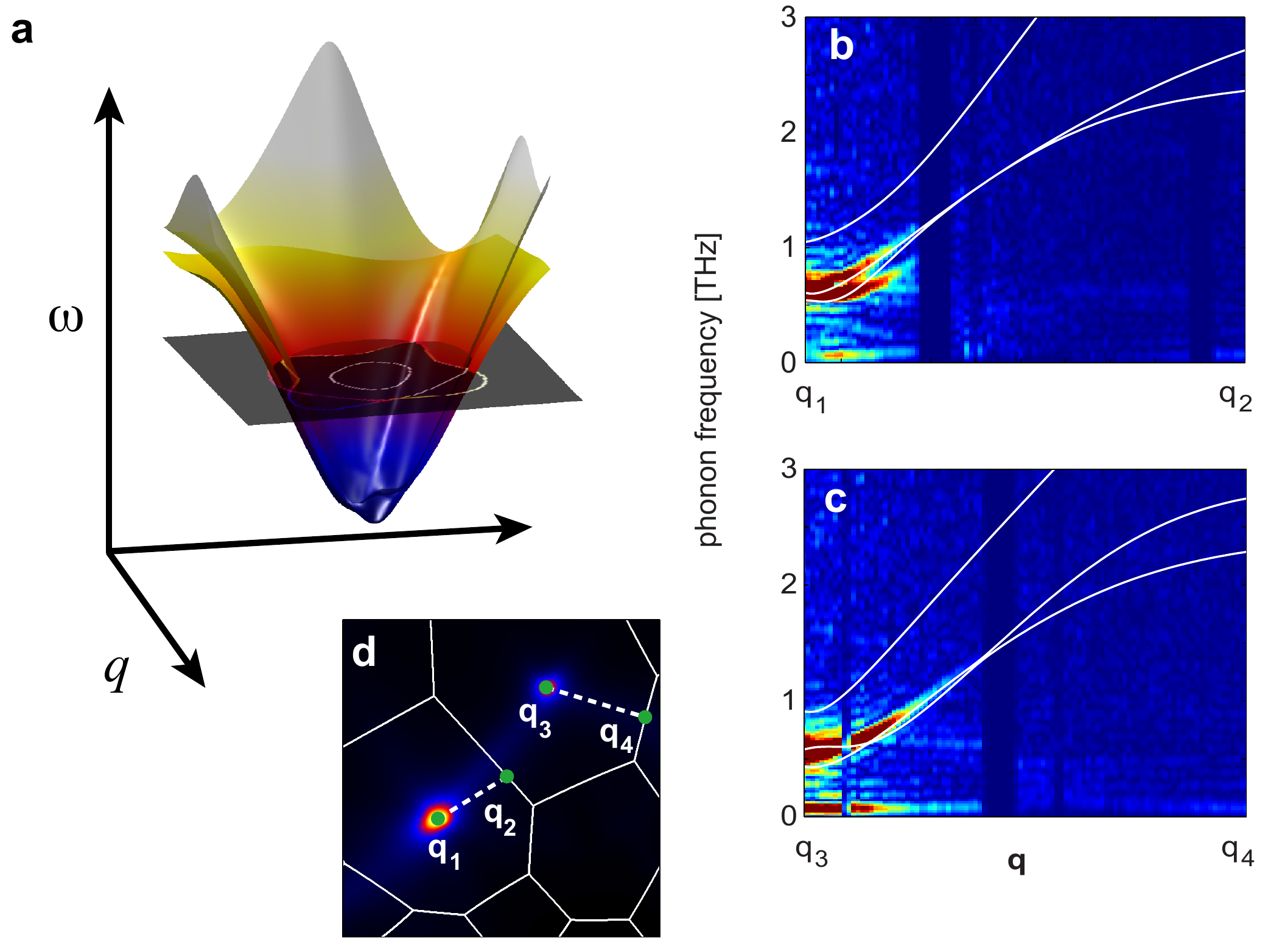}%
\caption{ {\bf Extracted dispersion relation in selected directions. }(a) Schematic representation of the constant-frequency cuts of the acoustic dispersion relation that yield the data in Fig.~\ref{Fig3}. The surfaces represent the two TA branches and the  plane represents a constant-frequency cut at $2\omega = 2.5$~THz. (b) and (c) acoustic dispersion along the sections shown with dashed lines in (d) where $q_1 = $\, [-0.09 -0.04 -0.07], $q_2 =$\,[-0.04 -0.82 0.37], $q_3=$\, [0.14 0.01 0.05] and $q_4 = $\,[0.27 -0.95 -0.07] (r.l.u). White lines in (b) and (c) represent the calculated acoustic dispersion. 
\label{Fig4}}
\end{figure}

Figure~\ref{Fig3} shows a zoomed view at selected frequencies of the Fourier transform (FT) along the time axis of the oscillatory component in Fig. \ref{Fig2} (top and bottom rows represent regions in boxes ``1'' and ``2'' in Figs.~\ref{Fig1} (a), respectively).
The intensity at each pixel and a given frequency is the magnitude of the FT of traces like those shown in Fig.~\ref{Fig2} (b). The bright loops appear at locations in momentum space where the intensity oscillates at the same frequency.
These contours (Fig.~\ref{Fig3}) represent constant-frequency cuts of the phonon dispersion relation as depicted schematically in Fig.~\ref{Fig4} (a). The data in Fig.~\ref{Fig3} show two bands, seen more clearly in the bottom row plots, which correspond to the two TA branches, with pinch points  where the bands are degenerate along high-symmetry directions. 
Their intensity depends on the amplitude of the coherent mean squared displacements, as well as their projection along $Q$.

Finally, Figs.~\ref{Fig4} (b) and (c) show the extracted dispersion relation along the directions indicated by the dashed lines in Fig.~\ref{Fig4} (d). Here the phonon frequency is $\Omega \approx \Omega' = \omega/2$ according to Eq.~(\ref{sqeezed_phonon}). It is important to note that we have not relied on any theoretical model of the interatomic forces to extract the phonon frequencies, only a knowledge of the scattering geometry and the time-delay is required.
The white lines in Figs.~\ref{Fig4} (b) and (c) are the calculated equilibrium dispersion. Note that within our experimental sensitivity we only pick out the transverse acoustic phonon branches and not the longitudinal acoustic branch. (currently the resolution is not high enough to resolve either optical phonon branch). This is to be expected as the excitation of carriers reduces the strength of the covalent bonds which give rise to the shear stability in the tetrahedrally bonded semiconductors\cite{martin1969}. Otherwise, the discrepancies are small and could be due to systematic errors in determining the sample orientation or the forces as much as changes in the excited state forces.
The curvature of the branches is due to our particular geometry, which results in a non-planar section of reciprocal space and avoids the Bragg peak at $q=0$ where $\Omega = 0$.
The flat spectral components at lower frequencies are likely due to fluctuations of the FEL that were not removed by our background subtraction. 
The sample was oriented far from the zone-center ($q=0$) to avoid strong Bragg reflections on the detector, particularly given the large wavelength fluctuations of the FEL.
Recently demonstrated self-seeded operation of the FEL\cite{amann2012} will provide better pulse stability with narrower bandwidth and will yield better momentum resolution, lower noise, and will allow sampling closer to $q=0$. An improved timing diagnostic\cite{coffee2013} will enable the observation of faster oscillations and thus higher frequencies.

The induced temporal coherences in the density-density correlations observed here are a  consequence of a sudden change in the interatomic potential and can thus be generalized to other excitations. These coherences span the entire Brillouin zone but will be favored in regions where the resultant (real or virtual) charge-density couples strongly to the phonons. For example it will be particularly strong in regions of enhanced electron-phonon coupling  and could find broad use in the study of the coupled degrees of freedom in complex materials. We further stress that, far-from equilibrium the pump-probe approach gives unique access to the phonon excitations and their interactions in the short-lived transient state.

%
%

This work was primarily supported by the US Department of Energy, Office of Basic Energy Sciences through the Division of Materials Sciences and Engineering under contract DE-AC02-76SF00515.
M.F. gratefully acknowledges financial support from the Volkswagen Foundation.
FQ and KST acknowledge support by the German Research Council (DFG) through the Collaborative Research Center 616 "Energy Dissipation at Surfaces".
J. L. was supported by the Swedish Science Council (VR)
A. H. was supported by AWE.  J. S. W.  is grateful for support from the UK EPSRC under grant no. EP/H035877/1.

\end{document}